# Measuring Long-Lived $^{13}$C-Singlet State Lifetimes at Natural Abundance

Kevin Claytor[1], Thomas Theis[2], Yesu Feng[2], and Warren Warren*[3]

[1]Department of Physics, Duke University, Durham, NC 27708, USA
[2]Department of Chemistry, Duke University, Durham, NC 27708, USA
[3] Departments of Chemistry, Radiology, Biomedical Engineering, and Physics, Duke University, Durham, NC 27708, USA

*warren.warren@duke.edu

Long-lived singlet states hold the potential to drastically extend the lifetime of hyperpolarization in molecular tracers for in-vivo magnetic resonance imaging (MRI). Such long lived hyperpolarization can be used for elucidation of fundamental metabolic pathways, early diagnosis, and optimization of clinical tests for new medication. All previous measurements of $^{13}$C singlet state lifetimes rely on costly and time consuming syntheses of $^{13}$C labeled compounds. Here we show that it is possible to determine $^{13}$C singlet state lifetimes by detecting the naturally abundant doubly-labeled species. This approach allows for rapid and low cost screening of potential molecular biomarkers bearing long-lived singlet states.

**Introduction**

Hyperpolarization methods can increase nuclear polarization by several orders of magnitude [1], allowing for the detection of MR signal from small amounts of material. One application of this technique is *in vivo* imaging of molecular tracers. For instance, pyruvate has been used to track cancer progression in mice [2] and more recently has been applied to human prostate cancer [3]. With current hyperpolarization methods such as dynamic nuclear polarization (DNP) [1] and para-hydrogen induced polarization (PHIP) [4–8] a wide variety of molecules can be hyperpolarized. Even with the dramatic increase in signal, the time for which these markers can be tracked is still limited by the spin-lattice relaxation time, $T_1$. For a long-lived nucleus such as $^{13}$C, this may be only tens of seconds. This restricts the biological processes that can be investigated to those that occur fast relative to $T_1$.

A solution to this lifetime challenge presents itself in long-lived singlet states [9–14]. The singlet state is an anti-symmetric combination of scalar coupled nuclear spins. Their symmetry properties make them resistant to several relaxation mechanisms [15,16] giving them longer lifetimes. For the same reason, singlet states are generally disconnected from the MR-probed part of the spin system. Despite this, a variety of techniques have been developed to access them. The singlet-triplet interconversion can be achieved by field-cycling [9,17], carefully designed pulse sequences such as magnetization-to-singlet (M2S) [12–14,18], chemistry [19], and more recently by low-powered RF fields (Spin-Lock Induced Crossing, or SLIC) [20,21]. Composite and adiabatic versions of SLIC [21] are able to generate the singlet over a wide range of field inhomogeneity and frequency offsets.

Many long-lived agents take advantage of lifetime gains by using a pair of low-$\gamma$ nuclei, such as $^{13}$C. Since $^{13}$C has a natural abundance of only 1.1%, singlet lifetime measurements have required the synthesis of doubly-labeled compounds, which is costly and time-consuming. This is particularly challenging because not all singlet states are long-lived. Pure theoretical modeling to identify structures with long-lived states is difficult because of the wide variety of relaxation mechanisms such as dipolar relaxation, chemical shift anisotropy, and spin-rotation interactions. Here we show that, for $^{13}$C, singlet-state lifetimes can be measured on natural abundance compounds, which can greatly accelerate the identification of molecules bearing long-lived singlet states.

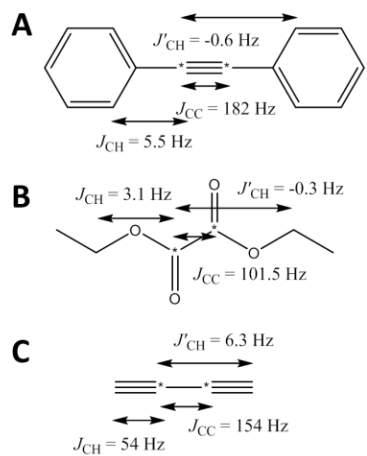

Figure 1: The molecules used in this study with the relevant J-couplings highlighted. A) DPA, B) DEO, C) Diacetylene.

We focus our attention on AA'X$_n$X'$_n$ systems and show proof of principle results for diphenyl-$^{13}$C$_2$-acetylene

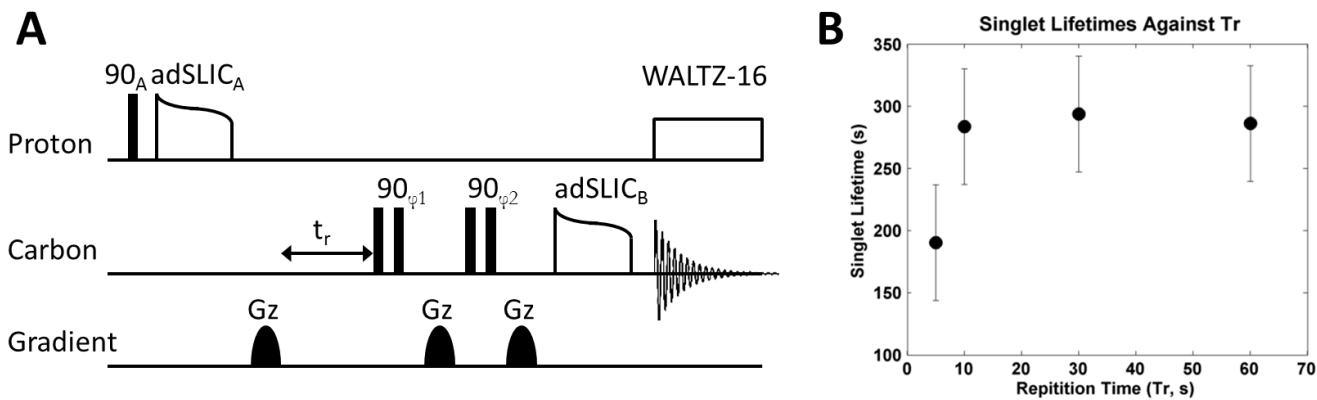

Figure 2: A) The proton-carbon adiabatic SLIC pulse sequence. A 12-component polyhedral phase cycle was used here with phase subscripts chosen to match the notation used in [24]. The two 90° pulses on the carbon channel are composite 90°'s ($90_\Phi 90_{\Phi+\pi/2}$). B) The singlet lifetime as a function of the repetition time for the labeled DPA sample. Although the singlet lifetime is several hundred seconds, a repetition time as short as ten seconds is able to accurately capture the singlet lifetime.

(DPA, n=2), diethyl-$^{13}C_2$-oxalate (DEO, n=2) and 2,3-$^{13}C_2$-diacetylene (n=1). These molecules and their couplings are shown in Figure 1. The idea of performing $^{13}C$-$^{13}C$ coherence experiments at natural abundance compounds is not new, underlying such techniques as INADEQUATE [22,23]. For example, with a cold probe and a sample at a high concentration at 700 MHz, we achieve an SNR of ~200 for the singly-labeled $^{13}C$ spectrum at natural abundance. The doubly labeled compound then would appear with an SNR ~1. Polarization stored in a singlet state is between 12% and 67% of this signal, so an SNR on the order of 10 can be achieved with a hundred averages, opening the possibility for an entire decay curve to be mapped overnight. The situation is more promising if the $^{13}C_2$-singlet polarization can be generated from proton polarization and/or transferred onto protons for detection, taking advantage of the higher proton gyromagnetic ratio [18].

Although there may be sufficient signal for detection, the thermal signal from the singly labeled species typically overlaps with the singlet signal and must be suppressed. The specific coherence pathways exploited by INADEQUATE do not work for singlet state detection, as they rely on a significant chemical shift difference between the coupled $^{13}C$ spins which is zero in the systems used here. Instead we suppress the thermal signal with filters composed of 90 degree pulses (since singlet states have no dipole moment, they are unaffected) followed by strong gradients, and with appropriate phase cycling [24].

For this to work, it is useful to measure as many molecular parameters as possible from the singly labeled species, and use sequences which can accept mismatch in the remaining parameters (such as the coupling $J_{CC}$ and the change in chemical shift upon double substitution with carbon-13). Exact knowledge of the carbon-carbon coupling value is critical to many of the approaches for pumping singlet states, such as the M2S pulse sequence. An alternative, spin-lock induced crossing with adiabatic shaped pulses [21] is less sensitive to the exact molecular parameters, surmounting this challenge and allowing singlet-state lifetimes to be measured from the natural abundance compound, even without exact knowledge of all parameters.

**Methods**

Experiments were performed using a 16.44 T (700 MHz) Bruker Ascend Spectrometer with a cryo-probe running Topspin 3.1. All experiments were temperature controlled to room temperature, except for diacetylene, which was held at 4° C. Neat samples were not used as a deuterium lock was needed to counteract the field drift during the experiments. The investigated samples were natural abundance diphenyl acetylene (DPA) (3.4 M), diethyl oxalate (DEO) (4.3 M), and diacetylene at (15 M), all in CDCl$_3$. Labeled diphenyl-$^{13}C_2$-acetylene (1.1 M) in CDCl$_3$ and $^{13}C_2$-diethyl oxalate (0.68 M) in DMSO were also available and measured as a standard for comparison to the natural abundance samples.

The pulse sequence is illustrated in Figure 2A. A hard 90° pulse excites the proton magnetization, which is followed by an adiabatic SLIC pulse on the proton channel to convert the thermal $^1H$-polarization into $^{13}C_2$-singlet polarization. As compared to a carbon-only experiment, this improves the signal by about $\gamma_H/\gamma_C \approx 4$ [18]. The amplitude of the adiabatic pulses is shaped like a tangent function with an offset [21];

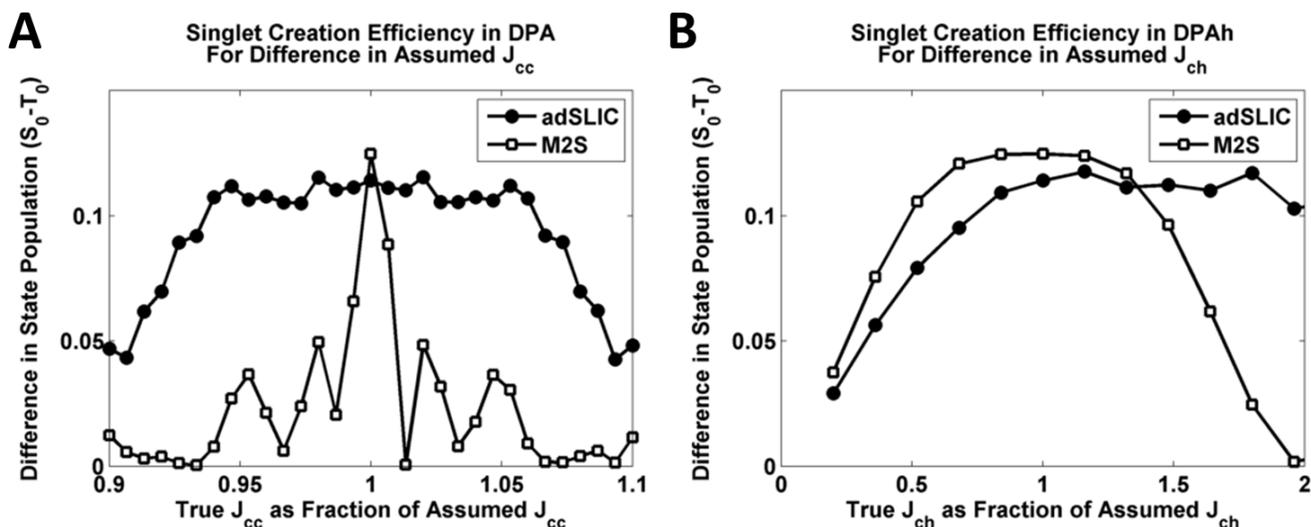

Figure 3: A) Simulations of the adiabatic SLIC (adSLIC) and M2S sensitivity to the carbon-carbon coupling constant. The plots show the population difference between the singlet and triplet states as the molecular $J_{CC}$ was varied from 90-110% (163.2 Hz-200 Hz) of the experimentally determined value of 181.8 Hz. The parameters of the adiabatic SLIC and M2S sequences were set using $J_{CC}$ = 181.8 Hz. While M2S slightly outperforms the adiabatic SLIC sequence on resonance, adiabatic SLIC still generates considerable singlet population even when the coupling value is substantially different from what is thought to be. B) A similar calculation showing the sensitivity as $\Delta = J_{CH} - J'_{CH}$ is varied. Simulations were performed using SPINACH [25].

$$\omega_1(t) = \omega_c - A\cos(\gamma)\tan(tA\sin(\gamma))$$
$$-\pi/(2A\sin(\gamma)) < t < \pi/(2A\sin(\gamma))$$

Equation 1

where $\gamma$ is an adjustable parameter ($\gamma=\pi/2$ is a flat pulse, $\gamma$ = 0.455 was used here), $\omega_c$ adjusts the offset and should be chosen as $\omega_c = 2\pi J_{CC}$, such that the shaped pulse adiabatically sweeps over the resonance condition of $\omega_1 = 2\pi J_{CC}$ at the center of the shape. Finally, the parameter A adjusts the maximum sweep excursion and sweep rate. This parameter is optimized depending on the difference in out-of-pair J-couplings, $\Delta = J_{CH} - J'_{CH}$, which ultimately drives the triplet-to-singlet interconversion. If $\Delta$ is small the sweep has to be slow and long (small A), whereas if $\Delta$ is large the sweep must occur over a wider range but can be faster (large A). For DEO ($\Delta$ = 3.4 Hz) we used A = 6.5, for DPA ($\Delta$ = 6.1 Hz) we chose A = 10.6, and for diacetylene ($\Delta$ = 47.7 Hz) we set A = 68. The corresponding pulse durations are 0.9 s for DEO, 0.5 s for DPA, and 0.075 s for diacetylene.

The key advantage of the adiabatic SLIC excitation, compared to a M2S sequence, is the broad excitation spectrum as a function of the difficult to determine carbon-carbon coupling, $J_{CC}$, shown in Figure 3A. With the wide excitation bandwidth of the adiabatic SLIC pulse the experiments can be performed successfully even with only a rough estimate for $J_{CC}$. The out-of-pair coupling $J_{CH}$ must also be known, but can frequently be read directly from the spectrum of the singly labeled compound. Even then, this coupling can be difficult to determine if it is small (below the spectral resolution of ~1Hz). Again, the adiabatic SLIC still performs robust singlet excitation without exact knowledge of that parameter, as demonstrated in Figure 3B.

Following the shaped SLIC pulse, the gradients before and after the singlet relaxation time ($\tau_r$) as well as the composite 90° pulses act as a filter for the thermal signal from the singly labeled species. To further eliminate signal from the singly labeled molecules, the sequence was phase cycled using a 12-component polyhedral phase cycle designed to pass the singlet signal [24]. The A, $\varphi_1$, $\varphi_2$, B notation in Figure 2A follows that of [24]. The singlet state is often characterized by a unique spectral signature which verifies that the sequence was generating the singlet state. After verification of the expected spectral signature, WALTZ-16 $^1$H decoupling was applied to improve the SNR when measuring the singlet decay.

Although the singlet state may be long-lived, it is always transformed back to magnetization for read-out after $\tau_r$, meaning that the relaxation delay between individual experiments, $T_R$, can be much shorter than the singlet lifetime. This decreases the total time required to obtain sufficient averages for a singlet decay curve with adequate SNR. Additionally, the polarization source in these experiments is from the proton polarization, with $T_1$ times that are even shorter than those of $^{13}$C, further reducing the necessary recycle delay. As shown in Figure 2B with the labeled DPA sample, an accurate lifetime can be determined with a repetition time of only 30 s (which was used for all experiments).

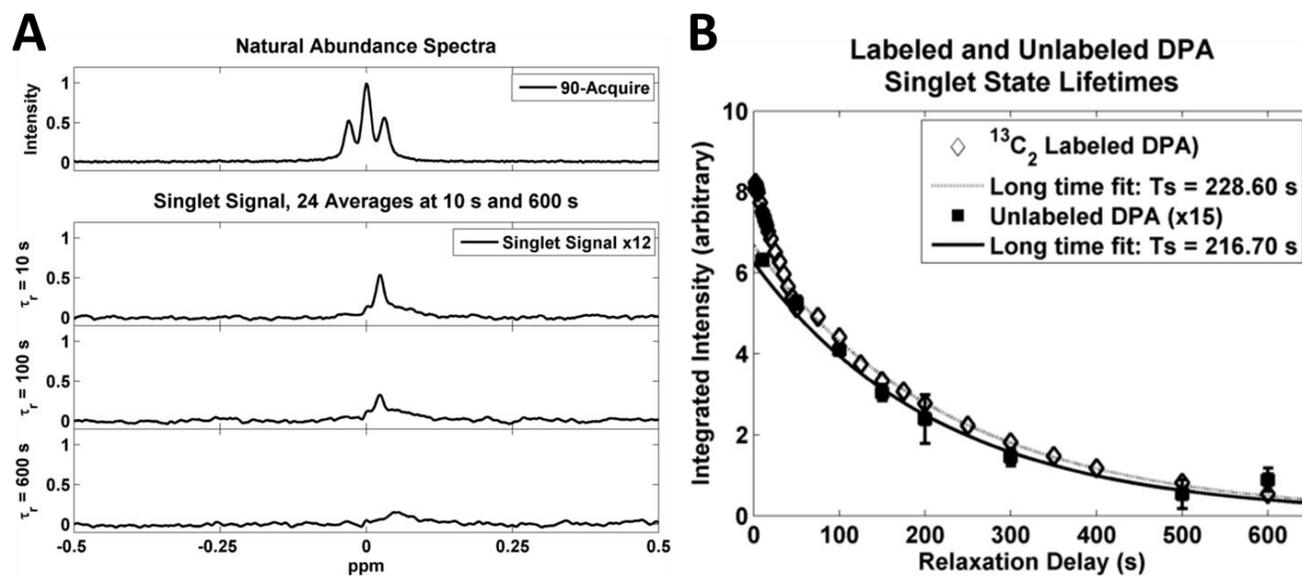

Figure 4: A) Spectra for diphenyl acetylene (DPA). Top shows the 90-acquire spectrum (single shot) of the alkyne carbons. The bottom spectra are for two different relaxation times (10, 100, and 600 s), see text for processing details. Decoupling was used, removing the structure of the singlet peaks, but improving SNR for the lifetime measurement. The 90-gradient sequence suppresses the thermal signal in favor of the singlet, which is essential because they overlap despite a small absolute chemical shift difference. B) Integrating over the singlet peak gives a singlet lifetime of ~220 s. This lifetime is in agreement with the corresponding experiment on the $^{13}$C$_2$-labeled compound.

The experimental data were adjusted by removing the first 30 ms of the FID, removing any low frequency noise. This was followed by line broadening of 1.5 Hz. The resulting spectra from DPA are shown in Figure 4A for a range of relaxation times.

**Discussions**

Figure 4 shows the experimental results for $^{13}$C$_2$-DPA. Part A shows the 90-acquire spectrum, dominated by the singly labeled species, and the signal from the doubly labeled species acquired after the pulse sequence is depicted below with relaxation delays of 10, 100, and 600 s. The signal decays to the baseline at long times indicating that there is no, or only a very small, leak-through of the thermal signal. Comparison of the natural abundance spectrum with the singlet signals also reveals a small absolute chemical shift difference for the doubly labeled species. Figure 4B shows the decay curve of the $^{13}$C$_2$-DPA singlet measured at natural abundance compared to the decay curve measured on the labeled compound. For the labeled compound it is obvious that the decay has a least two distinct exponential time constants; a rapid triplet state decay and a slow decay of the singlet state. While the natural abundance measurement does not have the time resolution to resolve the triplet and singlet decays separately (because of the required acquisition time), fitting of the long time component (t > 50 s) does give an accurate value for the singlet lifetime. All measured relaxation time constants ($T_1$ and Ts) may be found in Table 1.

The lifetimes observed for DEO at natural abundance and with the $^{13}$C enriched samples, Figure 5A, are very similar even in different solvents (CDCl$_3$ vs. DMSO) yielding a singlet state lifetime of 15 s. We previously measured a singlet state lifetime on DEO of 50 s at a field strength of 8.4T [14]. We believe that the larger chemical shift anisotropy of DEO gives rise to the faster relaxation time at the higher field. This is in contrast with DPA, which has a long singlet lifetime of 273 s at 8.4 T [18] and 216 s at 16.44 T, because of reduced CSA.

In the context of hyperpolarized biomolecular MRI, where the targeted field-strength is typically between 1 T and 7 T the field strength dependence of the relaxation times becomes very important. By measuring the singlet lifetime at a few field strengths, the relative contributions of CSA and dipole-dipole (DD) relaxation (typically the primary sources of relaxation) can be approximated, and the singlet lifetime predicted at lower fields. DD relaxation is roughly independent of the field strength, while CSA scales as the square of the field allowing us to construct a simple model for the singlet lifetime;

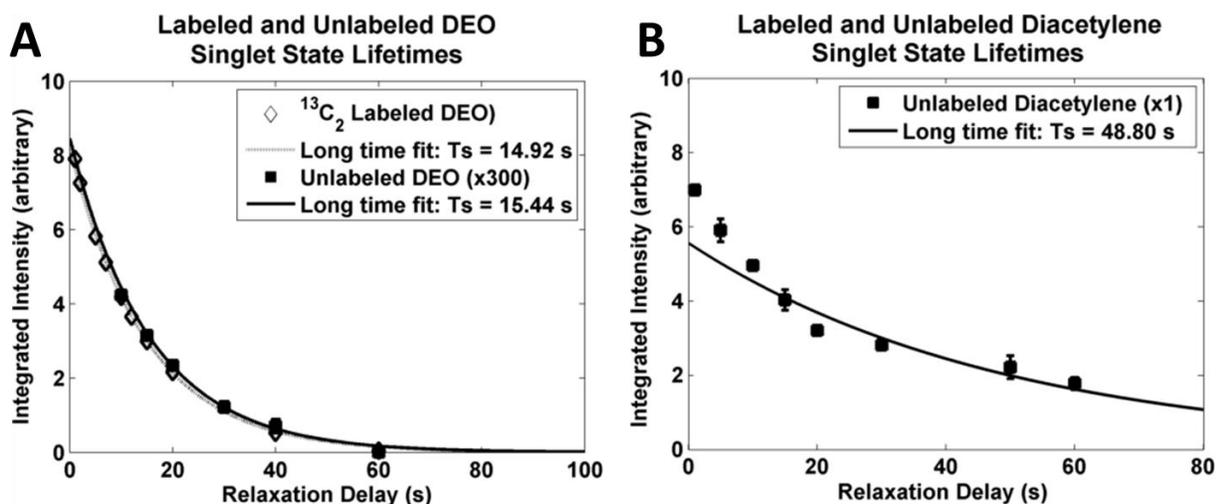

Figure 5: Singlet decay curves for A) DEO and the labeled DEO, and B) Diacetylene.

$$\frac{1}{T_S} = C_{DD} + C_{CSA} B_0^2 \qquad \text{Equation 2}$$

where $C_{DD}$ and $C_{CSA}$ indicate the relative contributions of the relaxation mechanisms to the singlet lifetime. Using the singlet lifetimes mentioned for DEO above, this would give a value of $T_S = 287$ s at 1 T. Of course this is only a rough estimate, simulations using density functional theory and molecular dynamics simulations and inclusion of all possible relaxation mechanisms would give a more accurate estimate of the singlet lifetime at lower fields.

With an accurate method of measuring singlet lifetimes at natural abundance, we now turn to diacetylene, where we do not have a corresponding doubly-labeled molecule. The singlet lifetime is shown with the $T_1$ time in Table 1, and the fit in Figure 5. Note that the $T_1$ times of the singly labeled species may be different from the $T_1$ for the doubly labeled species. Diacetylene is a particularly interesting example, as it is a prototypical AA'XX' system. Unlike DPA, diacetylene has fewer intramolecular couplings and at first glance one would suspect that it could have a very long lived singlet. The measured singlet state lifetime of ~40 s is, however, only 2x $T_1$. This is much less than the nearly 50-fold improvement on $T_1$ seen in DPA. We identify a few likely causes for the relatively short $T_S$. One is the very high concentration used, which promotes intermolecular relaxation processes. Another is the ratio of the carbon-carbon coupling ($J_{CC} = 154$ Hz) to the out-of-pair $J$-coupling difference ($\Delta = 47.7$ Hz). This is only ~3 for diacetylene and much smaller than for DPA ($J_{CC} / \Delta = 30$) or DEO ($J_{CC} / \Delta = 30$). This can lead to stronger coherent mixing effects between singlet and triplet states, but this effect could be offset through the use of decoupling. Finally, the field strength may play a role and the lifetime could be considerably longer at lower fields, and it will be interesting to examine the field strength dependence in future investigations.

Finally we point out that the singlet signal does not need to be well-separated from the spectra of the singly labeled species. For diacetylene, the peaks of the singly labeled species and those of the doubly labeled species do not overlap, giving us additional confidence in our measurements. The DPA and DEO spectra do not have this characteristic and there is overlap of the spectral features of the singly and doubly labeled species. Nonetheless, with the appropriate combination of filters and phase-cycling to suppress the thermal signal, the doubly-labeled singlet signal emerges, even though it is more than 200 times smaller, giving us reproducible singlet state lifetimes, as confirmed by comparison with the $^{13}C_2$ enriched compounds.

| Compound | $T_1$ (s) | $T_S$ (s) |
|---|---|---|
| DPA | 1.64 ± 0.02* | 216.70 ± 83.41 |
| DPA Labeled | 4.64 ± 0.07 | 228.60 ± 11.90 |
| DEO | 10.37 ± 0.16* | 15.44 ± 8.07 |
| DEO Labeled (in DMSO) | 8.17 ± 0.18 | 14.92 ± 0.56 |
| Diacetylene | 18.66 ± 1.16* | 48.80 ± 22.74 |

Table 1: $T_1$ values for the natural abundance compounds as well as their singlet lifetimes ($T_S$). Labeled versions of DPA and DEO were available and their $T_1$ and $T_S$ values are listed. The error listed is the 95% confidence interval of the fit. Stars indicate the $T_1$ value for the singly labeled compound.

## Conclusions

We have shown that it is possible to measure the singlet state lifetime of $^{13}C_2$-singlets at natural abundance of $^{13}C$, where $^{13}C_2$ pairs statistically only appear in $(1.1\%)^2$ of molecules. Currently, there are some limitations to this method that may be overcome in the future. The compounds analyzed here are liquids (DEO and diacetylene at 4°C) or highly-soluble solids (DPA in $CDCl_3$), so that highly concentrated samples can be prepared to provide sufficient SNR. A further increase of the SNR is achieved by employing a large magnetic field (16.44 T for the results shown). However, because of reduced CSA, singlet tracers often have longer lifetimes at lower magnetic fields. Extrapolation of the singlet lifetime to lower fields would be of great value because, to first approximation, SNR for body noise dominated hyperpolarized MRI is independent of the magnetic field strength. A rough possibility for extrapolating the singlet lifetime has been presented in Eq. 2. However, extrapolations can be made much more reliably by matching singlet state lifetimes and $T_1$ values measured at different field strengths (e.g. at 14.1 and 18.8T) to computational simulations which include all dominant relaxation mechanisms.

With the adiabatic SLIC sequence, we generate the singlet efficiently without requiring exact knowledge of the *J*-coupling parameters of the spin system. A rough estimate of the *J*-coupling parameters is sufficient to generate the singlet. This insensitivity to the exact parameters significantly eases the search to reveal the molecules with long singlet state lifetimes.

Being able to screen compounds at natural abundance avoids the costly and time consuming synthesis required to obtain the $^{13}C_2$ enriched compounds for the measurement of singlet state lifetimes. The case of diacetylene demonstrates this, which at first glance appears very promising for bearing a long-lived singlet state, but in fact does not, as proven by a measurement at natural abundance. Comparing singlet lifetimes to simulation will help to further refine singlet relaxation theory to make more accurate predictions. Just as importantly, this technique will allow for rapid identification of long-lived singlet compounds for use as tracers in biomolecular imaging applications.

## Acknowledgements

This work was supported by NSF under grant CHE-1058727.

[1] J.H. Ardenkjaer-Larsen, B. Fridlund, A. Gram, G. Hansson, L. Hansson, M.H. Lerche, et al., Increase in signal-to-noise ratio of > 10,000 times in liquid-state NMR., Proceedings of the National Academy of Sciences of the United States of America. 100 (2003) 10158–63.

[2] M.J. Albers, R. Bok, A.P. Chen, C.H. Cunningham, M.L. Zierhut, V.Y. Zhang, et al., Hyperpolarized 13C lactate, pyruvate, and alanine: noninvasive biomarkers for prostate cancer detection and grading., Cancer Research. 68 (2008) 8607–15.

[3] S.J. Nelson, J. Kurhanewicz, D.B. Vigneron, P.E.Z. Larson, A.L. Harzstark, M. Ferrone, et al., Metabolic Imaging of Patients with Prostate Cancer Using Hyperpolarized [1-13C]Pyruvate., Science Translational Medicine. 5 (2013) 198ra108.

[4] K. Golman, O. Axelsson, H. Jóhannesson, S. Månsson, C. Olofsson, J.S. Petersson, Parahydrogen-induced polarization in imaging: subsecond (13)C angiography., Magnetic Resonance in Medicine : Official Journal of the Society of Magnetic Resonance in Medicine / Society of Magnetic Resonance in Medicine. 46 (2001) 1–5.

[5] C. Bowers, D. Weitekamp, Transformation of symmetrization order to nuclear-spin magnetization by chemical reaction and nuclear magnetic resonance, Physical Review Letters. 57 (1986) 2645–2648.

[6] J. Natterer, J. Bargon, Parahydrogen induced polarization, Progress in Nuclear Magnetic Resonance Spectroscopy. 31 (1997) 293–315.

[7] S.B. Duckett, C.J. Sleigh, Applications of the parahydrogen phenomenon: a chemical perspective, Progress in Nuclear Magnetic Resonance Spectroscopy. 34 (1999) 71–92.

[8] C. Bowers, D. Weitekamp, Parahydrogen and synthesis allow dramatically enhanced nuclear alignment, Journal of the American Chemical …. (1987) 5541–5542.

[9] M. Carravetta, O. Johannessen, M. Levitt, Beyond the T1 Limit: Singlet Nuclear Spin States in Low Magnetic Fields, Physical Review Letters. 92 (2004) 153003.

[10] M. Carravetta, M.H. Levitt, Long-lived nuclear spin states in high-field solution NMR., Journal of the American Chemical Society. 126 (2004) 6228–9.

[11] G. Pileio, M. Concistrè, M. Carravetta, M.H. Levitt, Long-lived nuclear spin states in the solution NMR of four-spin systems., Journal of Magnetic Resonance (San Diego, Calif. : 1997). 182 (2006) 353–7.

[12] G. Pileio, M. Carravetta, M.H. Levitt, Storage of nuclear magnetization as long-lived singlet order in low magnetic field., Proceedings of the National Academy of Sciences of the United States of America. 107 (2010) 17135–9.

[13] M.C.D. Tayler, M.H. Levitt, Singlet nuclear magnetic resonance of nearly-equivalent spins., Physical

Chemistry Chemical Physics : PCCP. 13 (2011) 5556–60.
[14] Y. Feng, R. Davis, W. Warren, Accessing long-lived nuclear singlet states between chemically equivalent spins without breaking symmetry, Nature Physics. 8 (2012) 831–837.
[15] G. Pileio, Singlet state relaxation via intermolecular dipolar coupling., The Journal of Chemical Physics. 134 (2011) 214505.
[16] M. Carravetta, M.H. Levitt, Theory of long-lived nuclear spin states in solution nuclear magnetic resonance. I. Singlet states in low magnetic field., The Journal of Chemical Physics. 122 (2005) 214505.
[17] M.C.D. Tayler, I. Marco-Rius, M.I. Kettunen, K.M. Brindle, M.H. Levitt, G. Pileio, Direct enhancement of nuclear singlet order by dynamic nuclear polarization., Journal of the American Chemical Society. 134 (2012) 7668–71.
[18] Y. Feng, T. Theis, X. Liang, Q. Wang, P. Zhou, W.S. Warren, Storage of hydrogen spin polarization in long-lived 13C2 singlet order and implications for hyperpolarized magnetic resonance imaging., Journal of the American Chemical Society. 135 (2013) 9632–5.
[19] W. Warren, E. Jenista, R. Branca, X. Chen, Increasing hyperpolarized spin lifetimes through true singlet eigenstates, Science. 323 (2009) 1711–1714.
[20] S. DeVience, R. Walsworth, M. Rosen, Preparation of Nuclear Spin Singlet States using Spin-Lock Induced Crossing, arXiv Preprint arXiv:1307.0832. (2013).
[21] T. Theis, Y. Feng, T. Wu, W. Warren, Spin lock composite and shaped pulses for efficient and robust pumping of dark states in magnetic resonance, arXiv Preprint arXiv:1308.5666. (2013) 1–6.
[22] G.A. Morris, Sensitivity enhancement in nitrogen-15 NMR: polarization transfer using the INEPT pulse sequence, Journal of the American Chemical Society. 102 (1980) 428–429.
[23] A. Bax, R. Freeman, S.P. Kempsell, Natural abundance carbon-13-carbon-13 coupling observed via double-quantum coherence, Journal of the American Chemical Society. 102 (1980) 4849–4851.
[24] G. Pileio, M.H. Levitt, Isotropic filtering using polyhedral phase cycles: application to singlet state NMR., Journal of Magnetic Resonance (San Diego, Calif. : 1997). 191 (2008) 148–55.
[25] I. Kuprov, N. Wagner-Rundell, P.J. Hore, Bloch-Redfield-Wangsness theory engine implementation using symbolic processing software., Journal of Magnetic Resonance (San Diego, Calif. : 1997). 184 (2007) 196–206.